\documentclass[a4paper]{article}

\usepackage{INTERSPEECH2020}
\usepackage{graphbox}
\usepackage{color}

\title{Trust-UBA: A Corpus for the Study of the Manifestation of Trust in Speech}
\name{Lara Gauder$^{1,2}$, Pablo Riera$^2$, Leonardo Pepino$^{1,2}$, Silvina Brussino$^{3,4}$,\\Jazm\'in Vidal$^{1,2}$, Luciana Ferrer$^2$, Agust\'in Gravano$^{5}$}
\address{
$^1$Departamento de Computaci\'on, FCEyN, Universidad de Buenos Aires (UBA), Argentina\\
$^2$Instituto de Investigaci\'on en Ciencias de la Computaci\'on (ICC), CONICET-UBA, Argentina\\
$^3$Facultad de Psicolog\'ia, Universidad Nacional de C\'ordoba (UNC), Argentina\\
$^4$Instituto de Investigaciones Psicol\'ogicas, CONICET-UNC, Argentina\\
$^5$Escuela de Negocios, Universidad Torcuato Di Tella, Argentina}
\email{\{mgauder,priera,lpepino,jvidal,lferrer\}@dc.uba.ar;\\ silvina.brussino@unc.edu.ar; agravano@utdt.edu}

\begin{document}

\maketitle
\begin{abstract}
This paper describes a novel protocol for collecting speech data from subjects induced to have different degrees of trust in the skills of a conversational agent. The protocol consists of an interactive session where the subject is asked to respond to a series of factual questions with the help of a virtual assistant. In order to induce subjects to either trust or distrust the agent's skills, they are first informed that it was previously rated by other users as being either good or bad; subsequently, the agent answers the subjects' questions consistently to its alleged abilities. All interactions are speech-based, with subjects and agents communicating verbally, which allows the recording of speech produced under different trust conditions. We collected a speech corpus in Argentine Spanish using this protocol, which we are currently using to study the feasibility of predicting the degree of trust from speech. We find clear evidence that the protocol effectively succeeded in influencing subjects into the desired mental state of either trusting or distrusting the agent's skills, and present preliminary results of a perceptual study of the degree of trust performed by expert listeners. The collected speech dataset will be made publicly available once ready.
\end{abstract}

\section{Introduction}
The ability to dynamically monitor the user's mental state, including their engagement, satisfaction, and emotions in general \cite{kiseleva2016predicting,sano2016prediction,kiseleva2017evaluating} is becoming an increasingly important component of conversational agents. 
In particular, and especially for virtual assistants, tracking the user's degree of \emph{trust} in the system's skills may be critical for the success of the interaction. If a user starts displaying cues of distrust and the system can effectively detect such cues, then the dialogue manager could choose to act in consequence for regaining the user's trust. The main goal of this research project is the automatic detection of trust from speech, and in this paper, we focus on the collection of a speech trust dataset.

Several disciplines have investigated trust for decades, including psychology, anthropology, sociology, economics and political science. One important area of research has been the search for the factors that explain trust. Mayer et al.\ consider trust to depend on the trustor's perception of the trustee's \textit{ability, benevolence and integrity} \cite{mayer1995}. Other such factors have been proposed, including contextual and situational factors \cite{gulati1995,zucker1986}, and the propensity of a person to trust \cite{rotter1967}, among others. Trust has been also described as a dynamic phenomenon -- it can be created or destroyed during a conversation \cite{zand1972,korsgaard2014}.

Trust's nature has been depicted both as rational or cognitive \cite{coleman1990,hardin2006} and as emotional or affective \cite{miller1974}, or a combination of both \cite{mollering2006}. In either case, we hypothesize that the degree of trust affects and is affected by linguistic aspects (e.g.\ the form and content of discourse) and paralinguistic aspects (e.g.\ the intonation, pitch, speech rate and voice quality) of the trustor's and trustee's speech. The main goal of this research project is to study to what extent the trustor's degree of trust can be predicted from their speech signal using fully automatic methods.

To our knowledge, no speech corpus is currently available with annotations of varying degrees of trust, large enough to allow for statistical analyses and machine learning experiments. Consequently, we designed and implemented a novel protocol for collecting speech from subjects who are \textit{induced} to have different degrees of trust in the skills of a conversational agent. We subsequently used this protocol for building the \emph{Trust-UBA Database} in Argentine Spanish, which we describe in the present work. We also present preliminary results that indicate that  differences do exist in the speech produced under varying degrees of trust. Finally, we show that a team of expert listeners achieved very low agreement in the annotation of the perceived degree of trust based only on the speech signal, indicating the difficulty of the task.

\section{Protocol}
\label{sec:protocol}
The protocol consists of an interactive session in which the subject must respond to a series of questions, aided by a speech-enabled virtual assistant (VA). A text-only version of this protocol was first described and evaluated in \cite{gauder2019}. In this section we describe the speech-based version of the protocol in detail. 

\begin{figure}[!ht]
  \centering
  \fbox{\includegraphics[width=.95\linewidth]{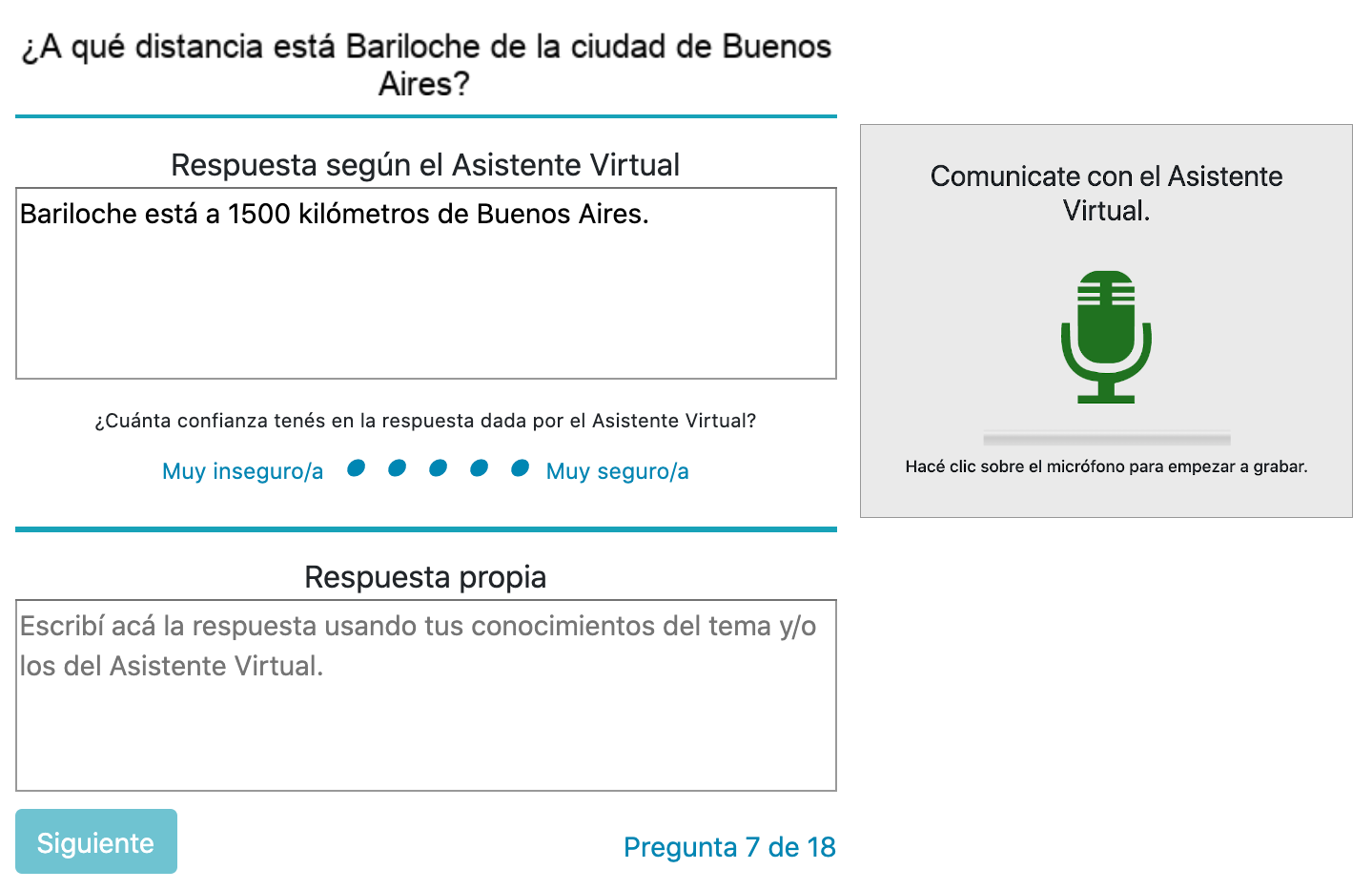}}
  \caption{Screenshot of the user interface for the question \textit{``What is the distance between Bariloche and Buenos Aires?''.} Subjects enter, from top to bottom: the response given by the VA, their confidence in the VA's response, and their own response. Subjects interact with the VA using the voice recording button shown on the right.}
  \label{fig:user-interface}
\end{figure}

\subsection{Session structure and initial bias}

The subjects' task is to respond a series of factual questions with the help of a VA. For each question, subjects are required to
\textbf{(1)} interact verbally with the VA to find the answer to the question;
\textbf{(2)} listen and transcribe the answer given by the VA;
\textbf{(3)} rate their confidence in the response given by the VA using a 5-level Likert scale; and 
\textbf{(4)} enter their own answer, based on what they believe to be correct (this may or may not match the VA's response). 
Figure \ref{fig:user-interface} shows a screenshot of the user interface: the current factual question is shown at the top left of the screen; below that is a form in which subjects must enter the VA's response, their confidence in the VA's response, and their own response. On the right lies a voice recording button used to communicate with the VA.

At the beginning of a series of questions, subjects are told that the VA they will interact with was previously rated by other users with either a very high or very low score: 4.9 and 1.4 out of 5 stars, respectively (these two values were chosen empirically based on pilots tests \cite{gauder2019}). These two conditions are central in our protocol and are meant to bias the user toward either trusting or distrusting the VA's skills. We refer to them as the \textit{high-score} and the \textit{low-score} conditions. 

With this setup we intend to benefit from a well-studied cognitive bias called \textit{anchoring} or \textit{previous-opinion} bias, in which a person's decision-making process is influenced by an initial piece of information offered to them, such as a house valuation made by another broker, or a patient diagnoses made by another doctor \cite{sackett1979bias,tversky1974judgment}.

Subsequently, the quality of the responses given by the VA is consistent with the informed abilities, making no mistakes in the high-score condition, and making some mistakes in the low-score condition. This is intended to reinforce in the subject the feeling that the former system is good, and the latter is bad.

\subsection{Types of factual questions}

Each series contains 18 factual questions, 6 of which we classify as \textit{easy} and 12 as \textit{difficult}. Easy questions are about topics that should be obviously known by anyone (e.g., \textit{``How many days are there in a week?''}) and are used to generate the feeling in the subject that the VA actually works. Difficult questions, on the other hand, were selected so that their correct answers would likely be unknown to most people (e.g., \textit{``What are the three longest rivers in Argentina?''}). Thus, for difficult questions subjects should depend on the VA's responses. Furthermore, from the subjects' perspective, difficult questions make the task more challenging and interesting; but from our part, these questions allow us to manipulate the subjects' varying degree of trust in the VA's skills.

Difficult questions may be answered either correctly or incorrectly by the VA, as a reinforcement of the corresponding initial bias presented to the subject. In the low-score condition, 6 of the 12 difficult questions are answered incorrectly; in the high-score condition, all 12 difficult questions are answered correctly.
Importantly, no easy questions are ever answered incorrectly, since we found in pilot tests that doing so typically caused unnecessary frustration in the subjects, along with an irreversible feeling that the VA is useless. For that reason, incorrect answers to difficult questions were chosen to trigger a sense of \textit{doubt} in the subjects; even though they may not know the correct answer, they should feel that the VA's answer is wrong, without seriously hurting its reputation. For example, for the question, \textit{``What is the distance between Barcelona and Madrid?''}, the VA's incorrect answer is 1000 km (it is actually 504 km).

In order to collect more speech data, questions can also be divided into two types, depending on the length of the interaction they are expected to trigger. Some questions and answers were prepared for forcing an exchange of several conversational turns. For example, after the subject asks \textit{``What is the melting temperature of aluminum?''}, the VA may ask what measurement unit it should provide the answer in (Celsius or Fahrenheit degrees). Using this strategy, we force subjects to have longer interactions with the VA and produce more dialogue acts -- not only questions but also answers. 

\subsection{Surveys}
\label{sec:evaluation-surveys}

At the beginning of each recording session, subjects are asked to complete a few surveys for demographic information (gender, age, birthplace, first and second languages), for personality type (15 dimensions mapped to the big five personality types \cite{mondak2010personality}), and for their degree of familiarity with, and trust in, virtual assistants and other digital systems.

\begin{figure}[ht]
  \centering
  \includegraphics[width=.75\linewidth]{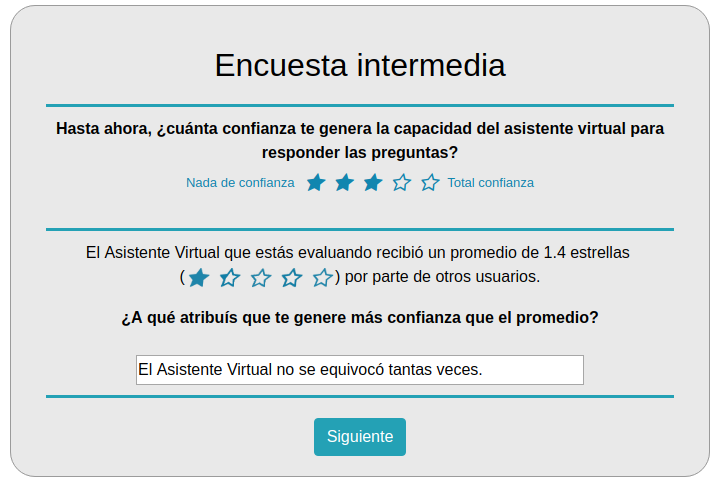}
  \caption{Screenshot of the evaluation survey.}
  \label{fig:va-survey}
\end{figure}

To assess the progress of the subjects' degree of trust throughout the series, they are required to complete simple \textit{evaluation surveys} after questions 6, 12 and 18. The first question, \textit{``So far, how confident are you in the system's ability to answer questions?,''}, is answered in a 5-level Likert scale presented using a 5-star metaphor, as seen in the top part of Figure \ref{fig:va-survey}. Only after answering this question, subjects are reminded that the current VA received an average of $X$ stars by other users (as explained above, $X=4.9$ in the high-score condition, and $X=1.4$ in the low-score condition), and are required to explain in a few words why their score was higher or lower than the average (bottom part of Figure \ref{fig:va-survey}). 

The purpose of the evaluation surveys is twofold. They measure the progression of the subjects' degree of trust along the series, and they also refresh the anchoring bias introduced at the series beginning, thus reinforcing the high- or low-score condition. The required textual explanation is intended to make subjects more conscious of this bias.

After completing a series of 18 questions and the third evaluation survey, subjects are required to rate how useful they found the VA, their degree of frustration with it, and how much they trusted it. They also report the extent to which they felt the following emotions and sentiments during the interaction:
active,
afflicted,
attentive,
tired,
decided,
disgusted,
distracted,
enthusiastic,
inspired,
uneasy,
nervous, and
fearful.
All questions are answered using a 5-point Likert scale.
These surveys are intended to further monitor and understand the subjects' behavior during their interaction with the VA.

\subsection{Implementation details}

Subjects interact with the VA via a voice recording button (right part of Figure \ref{fig:user-interface}), with which they may request the information needed to answer each question. The study interface was implemented online, to allow for data collection both in a controlled laboratory, and remotely over the Internet.

We built the VA dialogue system with the OpenDial toolkit \cite{lison2016opendial}, using a separate `dialogue domain' for each question -- i.e., a separate set of rules to trigger the system responses. We built a set of deterministic pattern-matching rules to cover the potentially many ways in which subjects may phrase their sentences.

We synthesized the VA's responses using Microsoft's publicly available speech synthesizer, with the HelenaRUS female voice in Spain Spanish with default settings.\footnote{https://azure.microsoft.com/en-us/services/cognitive-services/text-to-speech}
The subjects' utterances were transcribed using Google's publicly available automatic speech recognition system.\footnote{https://cloud.google.com/speech-to-text}

\begin{figure}[!t]
\centering
\includegraphics[align=c,width=.49\linewidth]{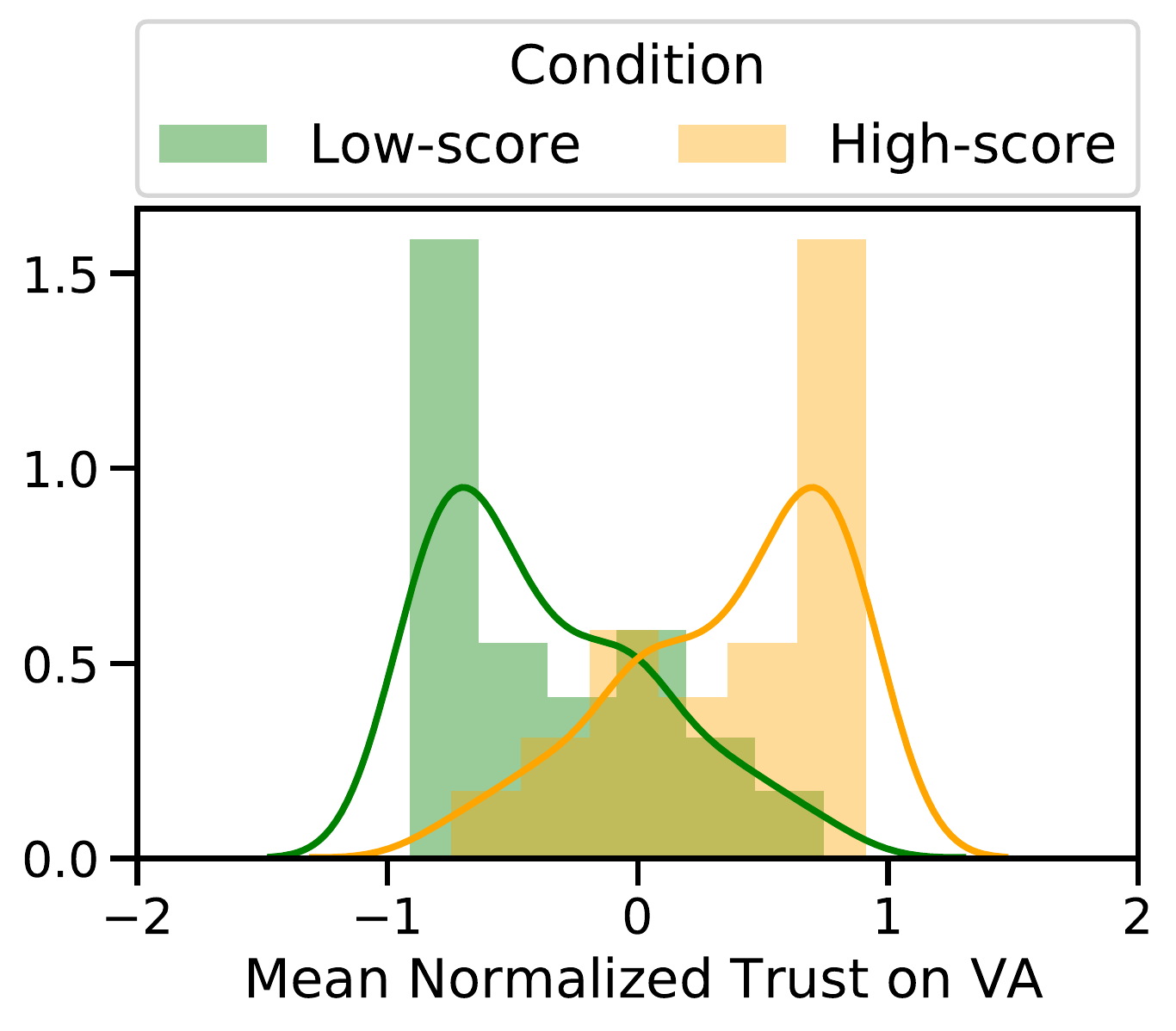}
\includegraphics[align=c,width=.49\linewidth]{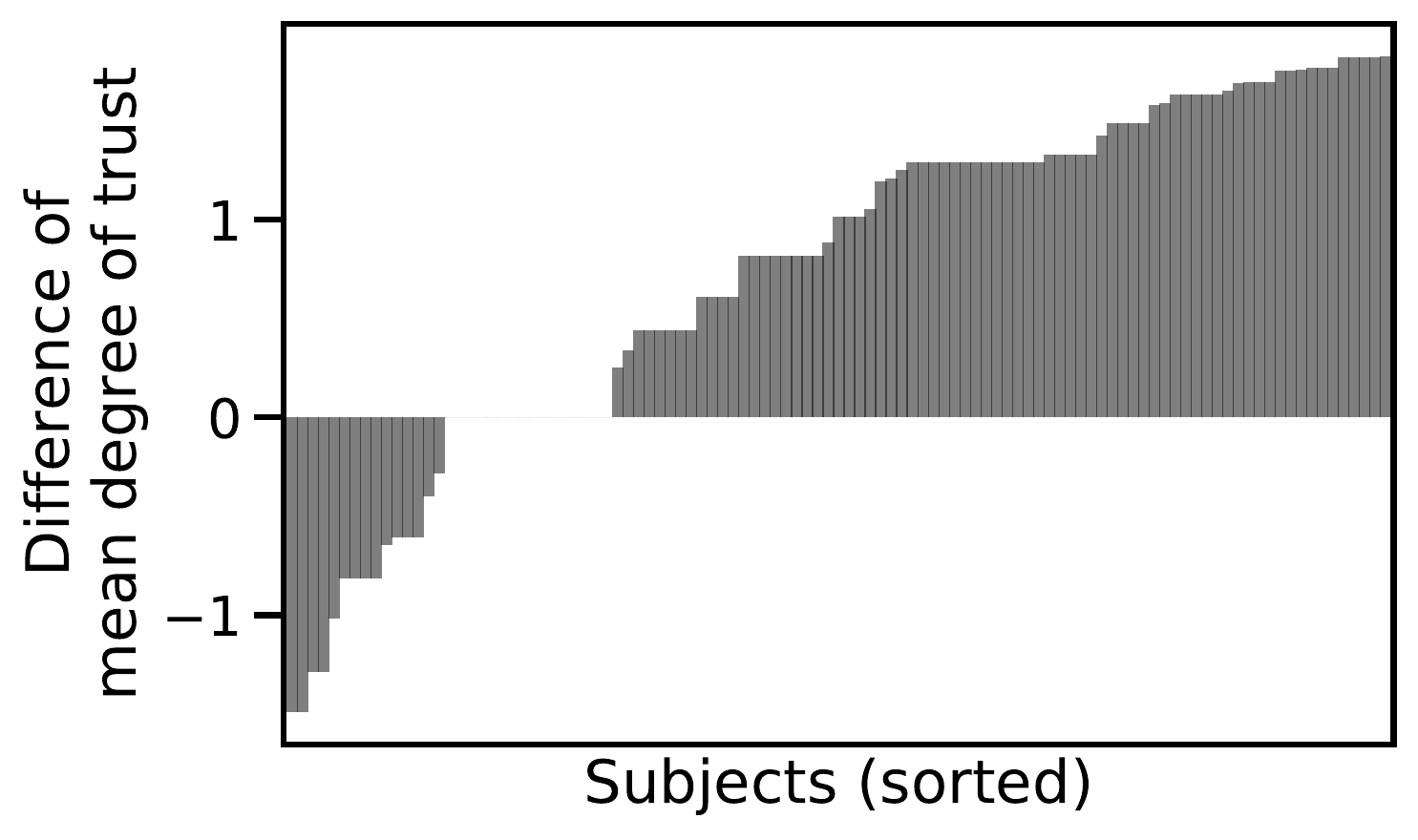}
\caption{Overall trust in the VAs' skills reported by subjects. Left: Histograms of means of normalized trust scores per condition. Right: Differences of mean trust scores from individual speakers across conditions.\vspace{-4mm}}
\label{fig:survey-dist}
\end{figure}

\section{Collected data}

Subjects were recruited via ads on social media, emails to student mailing lists, and posters at the University campus. 50 subjects participated at the University, in a controlled, silent environment (we call these the \textit{in-lab} subjects). This group was asked to solve two series of questions (one in each condition) and received a small monetary compensation for their time. The remaining 110 subjects participated over the Internet (these are the \textit{remote} subjects). In this case we had no control of the environment, which could result in poorer recording quality and lower concentration levels. This group was required to finish at least one series of questions, and were included in biweekly draws for a small monetary prize as compensation.

From the 160 volunteers, 83 were female, 76 were male, 1 did not reply. The mean age was 27.4 (stdev 9.2). Of these, 108 subjects completed two series of questions (50 in-lab, 58 remotely), one in each study condition (high- or low-score); the remaining 52 subjects (all remote) completed just a series in one condition.
In all, there are 131 and 137 completed series in the high- and low-score conditions, respectively. 
All subjects reported Spanish as their first language; all but 2 in-lab subjects, and all but 6 remote ones were born in Argentina. Thus, the collected speech is overwhelmingly in Argentine Spanish.

The collected data consists of 8493 short audios, with a mean duration of 4.7 seconds (stdev 2.1). Subjects that completed two series contributed on average with 62.4 audios (stdev 15.7); subjects that completed just one condition, 33.6 (stdev 8.1). 740 audios had to be excluded due to technical problems, such as network communication errors.

\begin{figure}[!t]
\centering
\includegraphics[align=c,width=.49\linewidth]{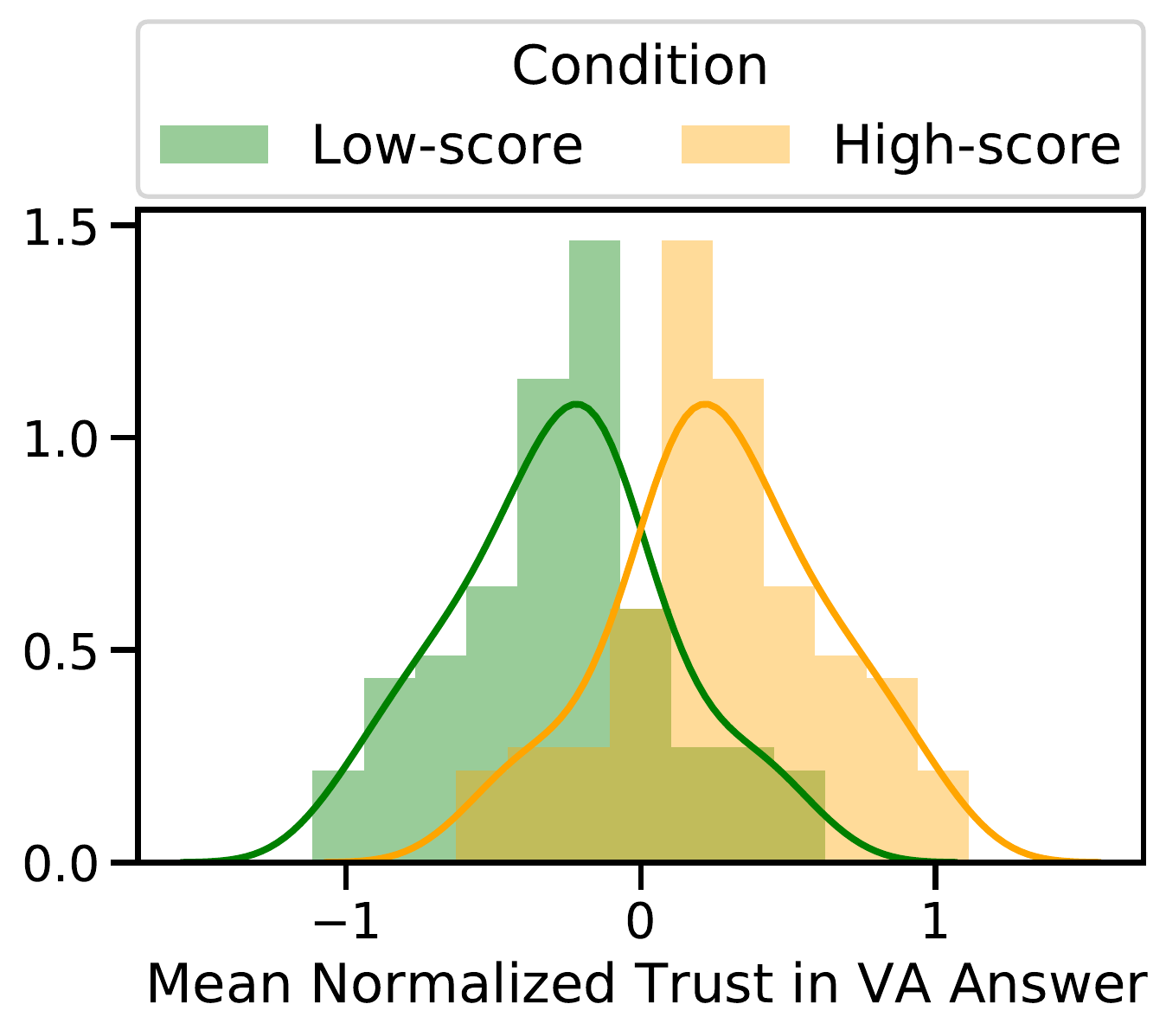}
\includegraphics[align=c,width=.49\linewidth]{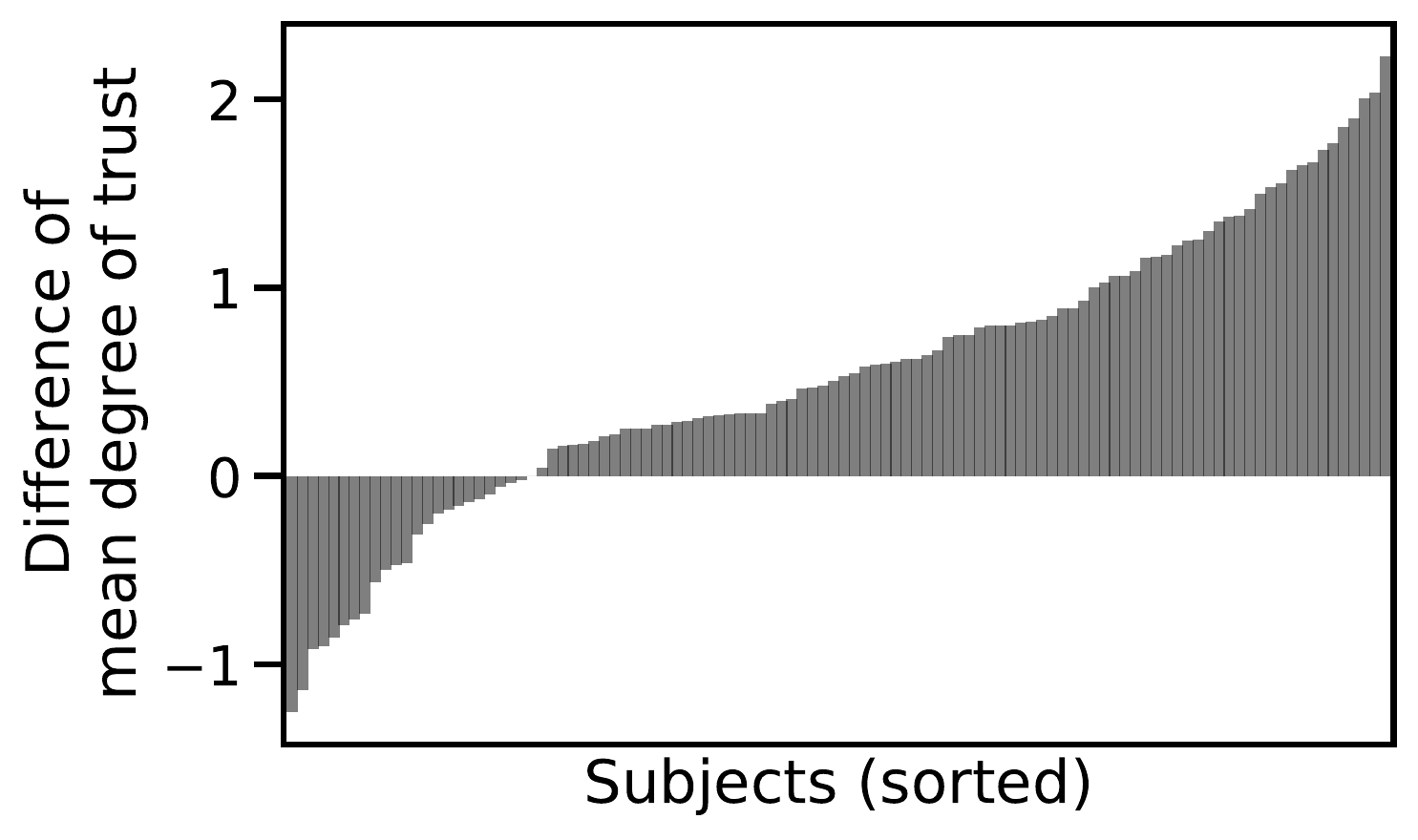}
\caption{Reported confidence in the responses given by the VAs to individual questions. Left:Histograms of means of normalized responses per condition. Right: Differences of mean responses from individual speakers across conditions.\vspace{-4mm}}
\label{fig:question-dist}
\end{figure}

\section{Protocol effectiveness}

The main purpose of the protocol is to induce subjects into either trusting or distrusting the VA's skills. In this section we analyze its effectiveness, by looking at  the subjects' responses to the evaluation surveys.  
The analysis that follows considers only the group of subjects that completed both study conditions: 50 in-lab and 56 remote subjects (2 remote subjects who completed both conditions were discarded because of transmission errors).

For the evaluations surveys, the score reported by subjects can be predicted with high significance ($p \approx 0$) by the condition it belongs to (low- or high-score). This was tested with a linear mixed model regression with effects given by subjects.

We subject-normalized all trust scores reported in the evaluation surveys (taken after questions 6, 12 and 18, see section \ref{sec:evaluation-surveys}), and computed the mean for each subject in each condition. Figure \ref{fig:survey-dist} (left) compares the distributions of such means in each condition. We observe a clear effect of condition type, with the low-score condition yielding significantly lower trust scores (linear mixed-effects model, $p \approx 0$).

Figure \ref{fig:survey-dist} (right) shows the differences across conditions of the (mean normalized) trust scores reported by individual subjects. In this plot, subjects are sorted to emphasize that only a few have an opposite effect (negative values) to what was intended by the protocol.

Likewise, figure \ref{fig:question-dist} shows similar plots, now with respect to the reported confidence in the VA's responses to difficult questions that were answered correctly by the VA (the most relevant ones for comparing the two study conditions). We observe clear differences in the distributions across conditions, as well as only a few subjects displaying an opposite effect (an increased trust in the low-trust condition). Based on this evidence, we conclude that the protocol succeeded in inducing the vast majority of subjects into either trusting or distrusting the VAs as intended.

\begin{figure*}[!t]
  \centering
  \includegraphics[width=1.0\linewidth]{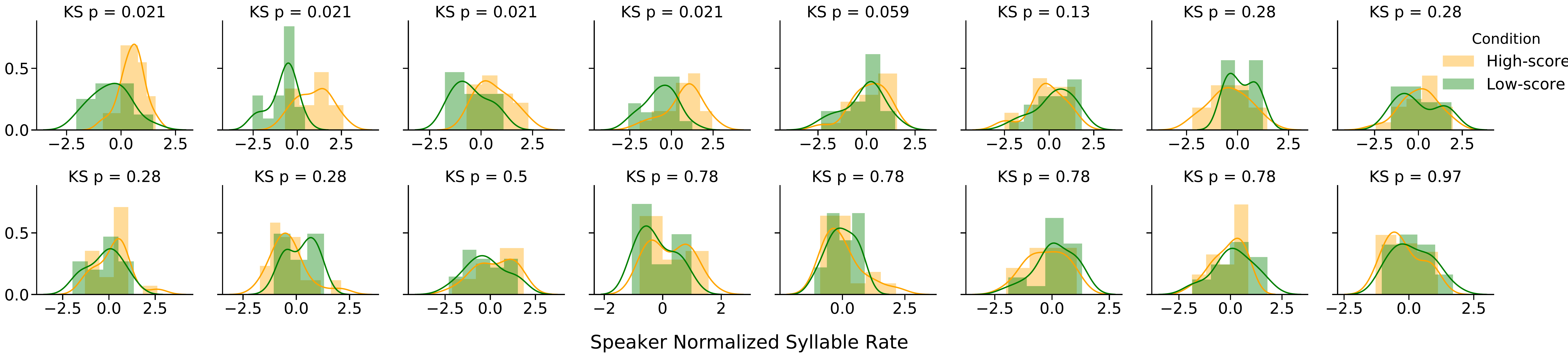}
  \caption{Comparison of the distributions of syllable rate without pauses for individual subjects in the high- and low-score conditions}
  \label{fig:sylrate}
\end{figure*}

\section{Preliminary study of the manifestation of trust in speech}

The main purpose of collecting this corpus is finding ways to automatically detect the degree of trust from speech. In this section we present the results of a preliminary study of speech rate, an acoustic/prosodic feature that, like hyper-articulation, characterizes speech directed at \emph{at-risk} listeners, such as infants or foreigners \cite{hazan2015,uther2007,saint2013motherese}. Speech directed at computers that make mistakes has been shown to have similar characteristics \cite{oviatt1998}. We thus hypothesize that speech rate may be a good predictor of the speaker's degree of trust in the VA's skills.

In figure \ref{fig:sylrate} we compare the syllable rate distributions for 16 individual subjects in the high-score and low-score conditions. (These were the in-lab subjects for which the system did not make speech recognition errors that might taint their trust levels. Therefore, these 16 subjects comprise the purest data points in our corpus.)
Syllable rates were computed from automatic alignments from hand corrected transcriptions. For each speaker, the distribution is composed of normalized mean syllable rates without pauses for every interaction with the VA. The $p$-values shown above each plot are from a Kolmogorov-Smirnov test, which measures how significantly different two distributions are. The four subjects with the more significant differences have a consistent shift across distributions with higher speech rate in the high-score distribution.

In conclusion, these preliminary findings hint the existence of a certain amount of variability in the syllable rate feature, relative to the study condition (low- vs.\ high-score VA). This variability suggests that the degree of trust may indeed be reflected in the speech signal. Our current experiments indicate that it may also be possible to automatically detect the degree of trust with machine learning models, using this and other acoustic/prosodic features \cite{pepino2020}.

\section{Perceptual annotation of trust}

An additional research question in the current project is whether humans are capable of telling solely from the speech signal whether the speaker trusted or distrusted the VA's skills. We first conducted informal pilot studies in which the authors tried to perform this task, only to find it extremely difficult. We thus decided to gather a team of psychology researchers and practitioners who, as experts in human behavior, would be good candidates for succeeding in this task.

As a result, ten female expert annotators were asked to listen to a pair of sequences of audios from each in-lab subject (they thus listened to 50 pairs of audios). Each such sequence was formed by the first recordings produced by the subject for each of the final six questions in a series -- during which we expect the trust/distrust effect to be at its maximum level. The six audios in a sequence were merged into a wav file, separated by a simple tone. 

Each pair of sequences was presented to annotators on a web page, in random order. For each pair, they had to select which audio corresponded to utterances directed at the less trustworthy VA, together with their confidence level in a 5-level Likert scale. 
At the end of this study, annotators were asked to write in their own words what factors they considered when conducting this task. All annotators were paid for this task.

We examined inter-annotator agreement using Fleiss' $\kappa$ measure \cite{Fleiss71}, which yielded a value of 0.116. This is interpreted as ``slight'' agreement above chance. We also conducted a permutation test, which confirmed that this slight agreement is indeed significantly not random ($p \approx 0$). This suggest that the annotators did perceive certain speech cues related to trust, albeit faint and unreliable ones. An analysis of their written reports does not reveal any clear consensus.

\section{Conclusions}

The main conclusion of this work is that the proposed protocol manages to influence participants into a particular mental state. In line with \cite{gauder2019}, we provide further evidence supporting the hypothesis that subjects effectively trusted more in the abilities of one of our two virtual assistants (VAs). Therefore, the collected speech corpus, the Trust-UBA Database, shall be a valuable contribution to the research community, as a useful first dataset for studying trust in speech.

As a preliminary proof of concept, we also presented evidence that an important prosodic/acoustic feature, speech rate, appears to show a significant amount of variation across the two study conditions (low- and high-score VAs). This is an promising first step towards the challenging task of automatically detecting the degree of trust from the speech signal using machine learning techniques.

Finally, we conducted a perceptual test, in which a group of Psychology researchers and practitioners rated the degree of trust by only listening to the audio samples. The resulting inter-annotator agreement was extremely low, but still above chance level, suggesting that even though the task is very difficult, they were still able to pick up some faint signal in the data.

\section{Acknowledgements}
This material is based upon work supported by the Air Force Office of Scientific Research under award no.~FA9550-18-1-0026. 
\bibliographystyle{IEEEtran}

\bibliography{mybib}

\end{document}